\documentclass{elsart}

\usepackage{natbib}
\usepackage{epsfig}

\usepackage{amssymb}

\begin{document}

\begin{frontmatter}

\title{The prompt gamma-ray emission of novae}

\author{M. Hernanz}
\address{Instituto de Ciencias del Espacio (CSIC) and 
Institut d'Estudis Espacials de Catalunya IEEC,
Edifici Nexus, C/Gran Capit\`a 2-4, E-08034 Barcelona, SPAIN}
\author{J. G\'omez-Gomar}
\address{IEEC}
\author{J. Jos\'e}
\address{IEEC and Departament de F\'{\i}sica i Enginyeria Nuclear (UPC), 
Avinguda V\'{\i}ctor Balaguer, s/n, E-08800 Vilanova i la Geltr\'u (Barcelona),
SPAIN}

\begin{abstract}
Classical novae are potential gamma-ray emitters, because of the 
disintegration of some radioactive nuclei synthesized during the explosion. 
Some short-lived isotopes (such as $^{13}$N and $^{18}$F), as well as the 
medium-lived $^{22}$Na, decay emitting positrons, which annihilate with 
electrons and thus 
are responsible for the prompt emission of gamma-rays from novae. 
This emission consists of a 511 keV line plus a continuum between 
20 and 511 keV, and is released before the maximum in visual luminosity, i.e., 
before the discovery of the nova. The main characteristics of this prompt 
emission, together with the related uncertainties (both of nuclear and 
hydrodynamical origin, with a particular emphasis on the influence of the 
envelope properties) and prospects for detectability are analyzed in this 
paper. 
\end{abstract}


\end{frontmatter}

\vspace{-0.5cm}
\section{Introduction and models}
\vspace{-0.5cm}
The explosive phenomenon of classical novae occurs on the surface of 
accreting white dwarfs, in close binary systems of the cataclysmic variable 
type, whenever certain initial conditions of white dwarf mass and luminosity 
and mass-accretion rate are met. The hydrogen-rich accreted matter, mixed 
with matter from the underlying white dwarf core, burns in degenerate 
conditions, leading to a thermonuclear runaway and the ensuing explosion.
An important increase in visual luminosity and the ejection of a fraction 
of the accreted envelope are some of the consequences of the explosion. The 
outburst is accompanied by the release of gamma-rays, with different time scales. 

The prompt $\gamma$-ray emission from classical novae has its origin in 
e$^+$-e$^-$ annihilation, with the positrons coming mainly from 
$^{13}$N and $^{18}$F decays (see the pioneering work from \citet{LC87}). 
The positrons emitted by $^{22}$Na 
decay also contribute to the prompt emission, but with a much lower flux than 
those from $^{13}$N and $^{18}$F, because of its much longer decay time. 
In general, the released 
positrons annihilate and produce a line at 511 keV and a continuum below it. 
The continuum is produced both by the positronium 
emission (when it is formed in triplet state) and by the Comptonization of 
the photons emitted in the line; it has a cut-off at around 20-30 keV, 
because of photoelectric absorption (this mechanism has larger cross-sections 
than Compton scattering at low energies). The isotopes $^{13}$N and $^{18}$F 
have relatively short lifetimes, which makes the corresponding emission of 
short duration and tightly related to the conditions in the expanding 
envelope (opacity to gamma-rays). The prompt gamma-rays from novae are emitted 
very early in the explosive phase, i.e., before the maximum in visual 
luminosity and, therefore, before optical discovery for the majority of novae. 

There is another type of gamma-ray emission, related to line emission from 
the decay of medium-lived radioactive nuclei (such as 478 keV from $^7$Be 
and 1275 keV from $^{22}$Na), lasting for months and years after the 
explosion, which has not been detected in any nova up to now \citep{Har91,
Har96,Lei88,Iyu95}, but its study is out of the scope of 
this paper (see for instance the pioneering papers from \citet{Cla81}, 
\citet{CH74}, and the recent ones from \citet{Gom98}, \citet{Her99a} 
and references therein).

Complete evolution of various nova models (both of the CO and ONe type, 
depending on the chemical composition of the underlying white dwarf, which, 
in turn, depends on its core mass) from the accretion phase up to the 
ejection one, has been computed, with a hydrodynamical code (see \citet{JH98}, 
for details about the code). It is important to stress that some nuclear 
reaction rates play a very important role on the synthesis of the most 
relevant radioactivities in nova explosions. In particular, $^{18}$F synthesis 
is affected mainly by $^{18}$F+p reactions ($^{18}$F(p,$\gamma$) and 
$^{18}$F(p,$\alpha$), which are still quite uncertain (see \citet{Her99b}
and \citet{Coc00}). Therefore, the final amount of $^{18}$F, which almost 
directly translates into the flux of positron annihilation gamma rays emitted 
promptly, will be as close to reality as the measured nuclear cross sections 
are. It is also important to treat accurately the initial composition of the 
underlying white dwarf core, specially for the case of the ONe novae, as 
well as the amount of mixing between accreted and white dwarf matter (see 
discussion in \citet{JH98}).   
In table \ref{radioac} we show the radioactivities, relevant for the positron 
annihilation gamma rays, for a handful of computed nova models. One important 
result is that similar amounts of $^{18}$F are produced in both nova types, 
contrary to what happens with other longer lived radioctivities (i.e., $^7$Be,
mainly produced in CO novae, and $^{22}$Na, mainly produced in ONe novae).  

\begin{figure}
\setlength{\unitlength}{1cm}
\begin{picture}(18,8)
\put(0,-2){\makebox(9,10){\epsfxsize=8cm \epsfbox{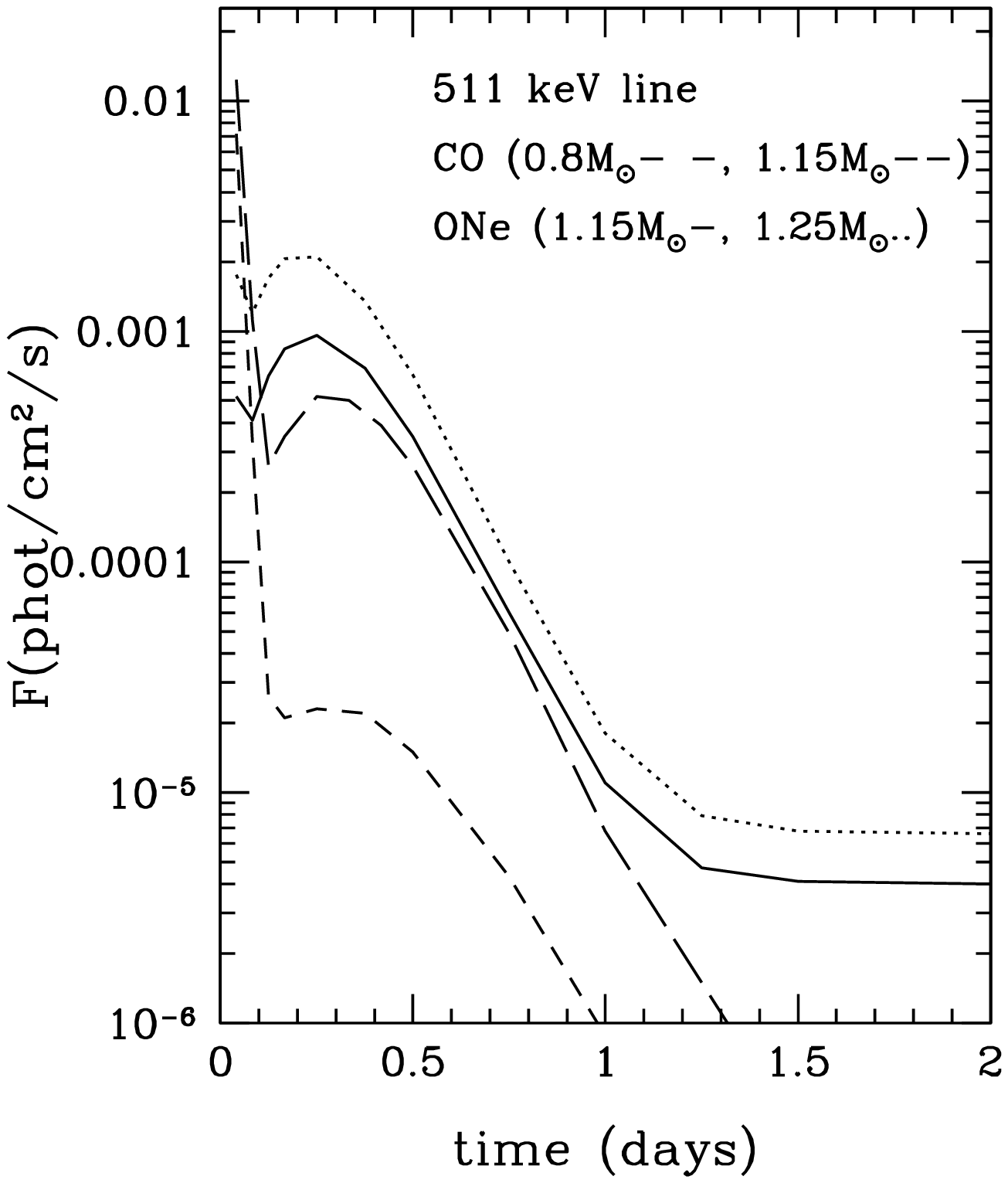}}}
\put(7.5,-2){\makebox(9,10){\epsfxsize=8cm \epsfbox{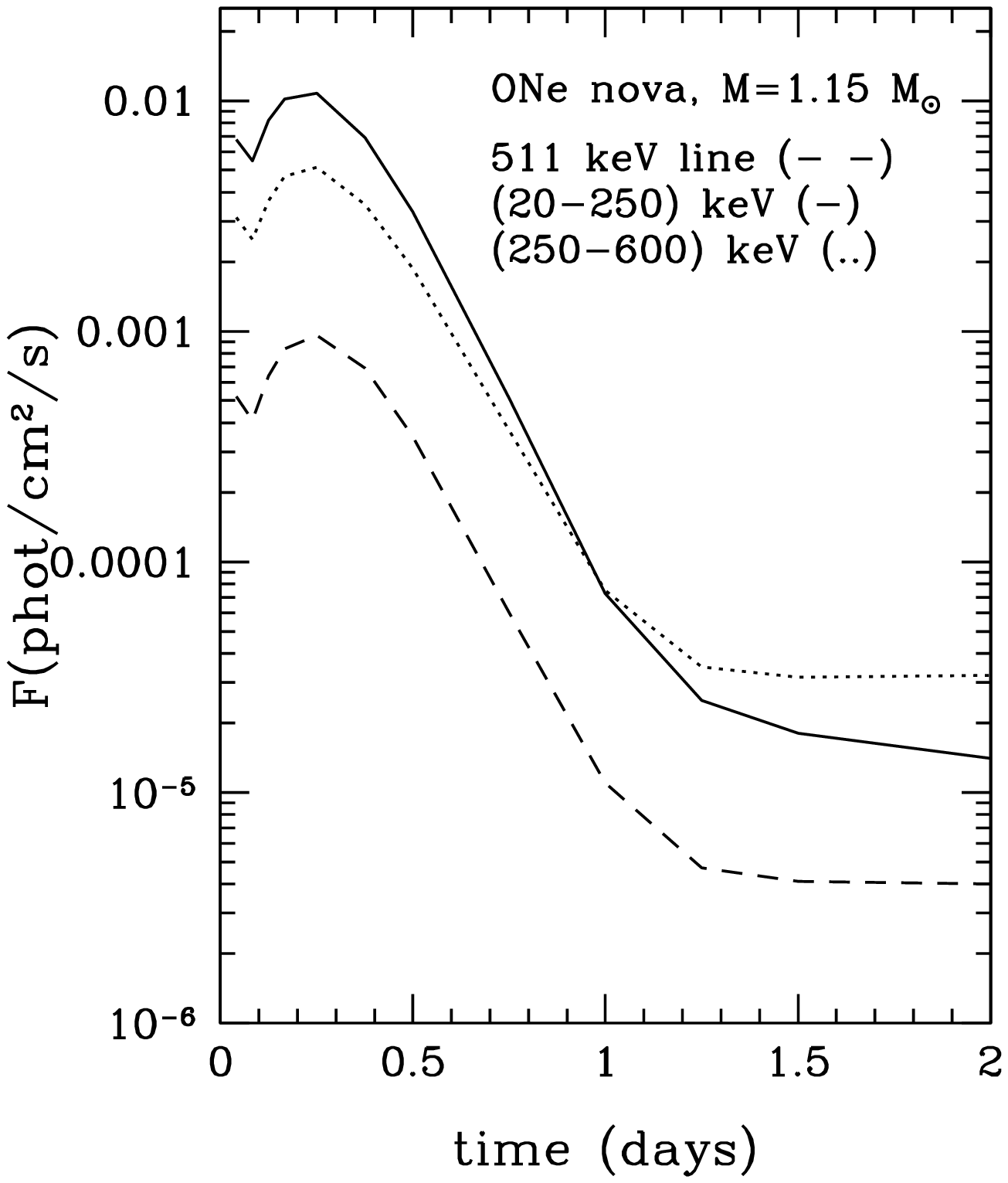}}}
\end{picture}
\caption{(Left) Light curves for the 511 keV line of the 4 nova models shown 
in table \ref{radioac}, placed at a distance of 1 kpc . 
(Right) Continuum light curves for the ONe nova of 1.15 M$_\odot$ at the same 
distance.}
\label{lcannihil}
\end{figure}

The main properties of the emission of positron annihilation gamma rays from 
novae are displayed in figure \ref{lcannihil}, where the time axis origin is 
at peak temperature (which occurs before visual luminosity maximum). First we 
show (figure 
\ref{lcannihil} left) the light curve of the 511 keV line, for the four models 
from table \ref{radioac}. Although the amount of ejected $^{18}$F is similar 
in all the models, the less massive CO nova (0.8 M$_\odot$) emits a smaller
flux than the other ones; that's because its outer expanding shells move at 
smaller velocities and, therefore, are more opaque to gamma-rays. It is important 
to notice that there is also an initial peak in the light curves, related to 
$^{13}$N decay, of even shorter duration than that produced by $^{18}$F decay 
(because of the shorter lifetime of $^{13}$N as compared with $^{18}$F). Another 
interesting aspect is that in ONe novae the 
511 keV line has a longer duration, because of the contribution of the positrons 
from $^{22}$Na-decay: a low-level ``plateau'' becomes visible at the tail of the 
light curve (see figure \ref{lcannihil}). 
However, when the envelopes finally become transparent (one or two weeks 
after T$_{\rm peak}$) e$^+$ escape without annihilating.
A similar behavior is displayed by the continuum light curves (see figure 
\ref{lcannihil} right); here one can see that there is much more flux in the 
continuum than in the 511 keV line (which has widths FWHM between 3 and 8 keV). 
\vspace{-0.5cm}
\section{Influence of the ejecta properties in the annihilation emission}
\vspace{-0.5cm}
Other important factors, beyond the critical nuclear reaction rates mentioned 
above, concern aspects of the envelope. As an illustrative example, the low 
mass CO nova (0.8 M$_\odot$) emits a flux in the 511 
keV line, which is smaller than that emitted by other more massive novae by a 
factor larger than their ratio of $^{18}$F yields (see figure 
\ref{lcannihil} left and table \ref{radioac}). In order to study these factors, 
we varied ejected masses and velocity profiles in some of our models, leaving 
the abundances of radioactive material unchanged. 
These models are interesting for illustrative purposes, although they are not 
self consistent. An additional interest comes from the fact that all current  
theoretical nova models still fail to reproduce some observed ejected masses 
(observed larger than theoretical).

\begin{figure}
\setlength{\unitlength}{1cm}
\begin{picture}(18,8)
\put(0,-2){\makebox(9,10){\epsfxsize=8cm \epsfbox{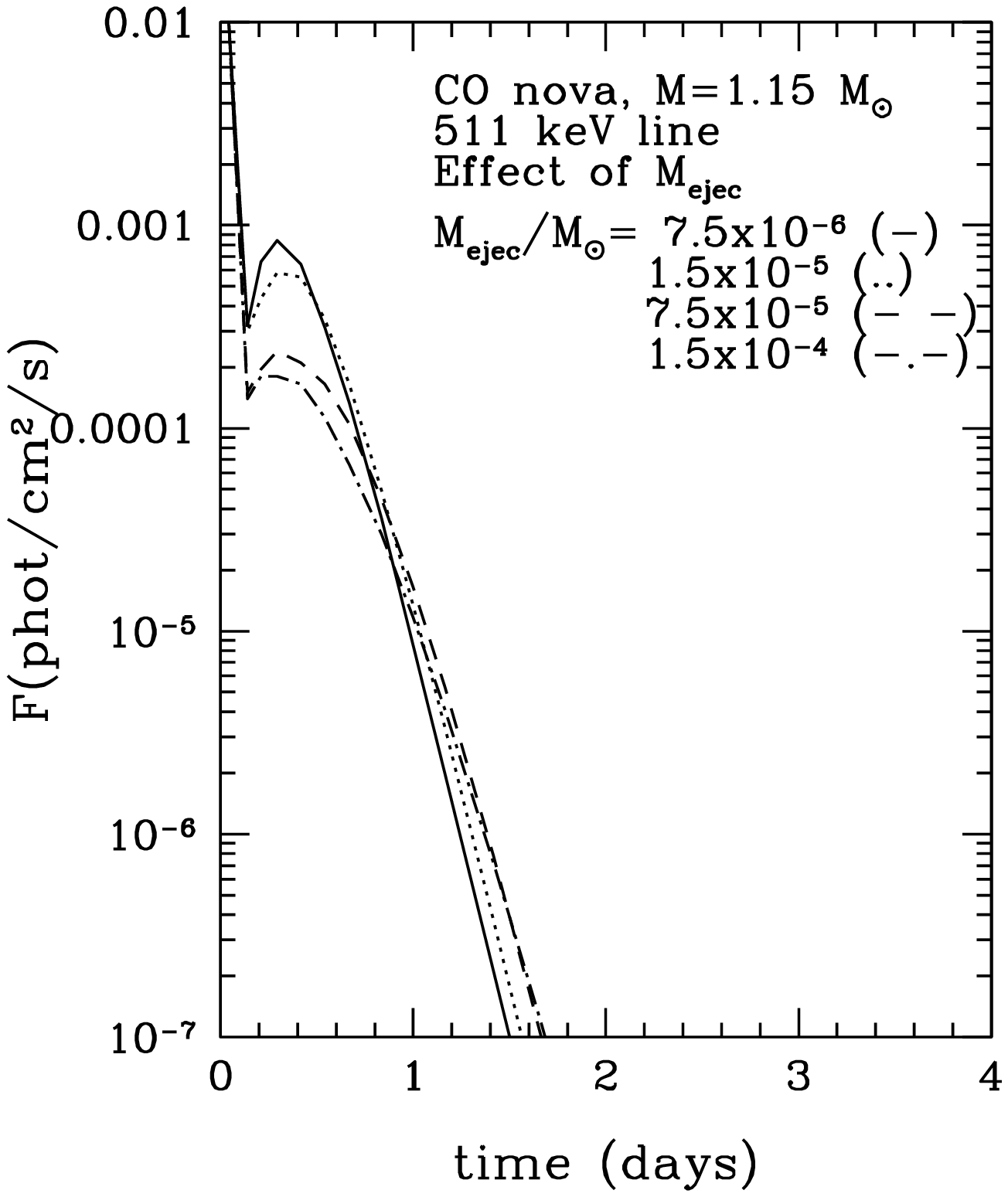}}}
\put(7.5,-2){\makebox(9,10){\epsfxsize=8cm \epsfbox{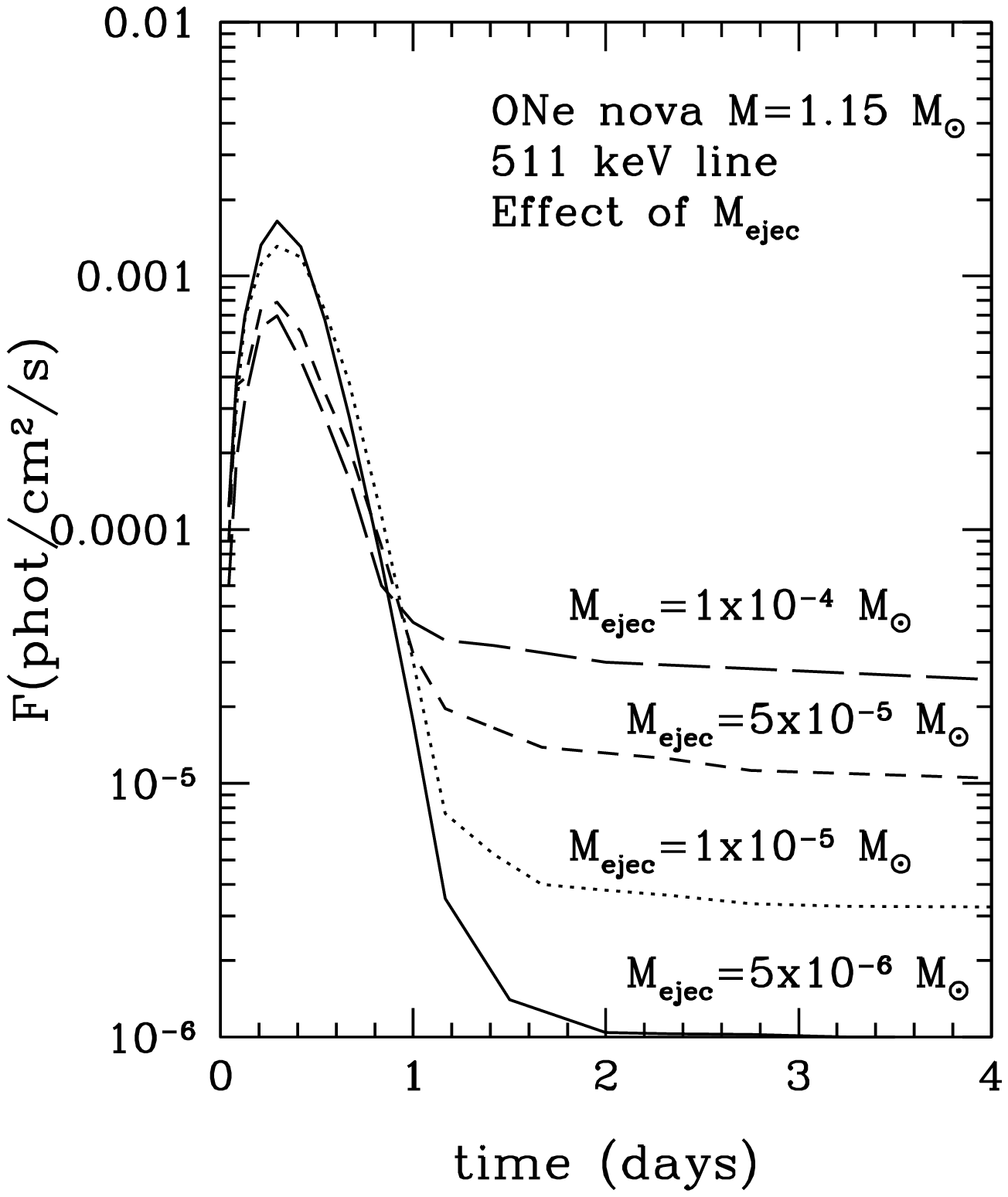}}}
\end{picture}
\caption{(Left) Light curves for the 511 keV line for a CO nova of 
1.15 M$_\odot$, for a range of ejected masses. (Right) Same for an ONe nova of 
1.15 M$_\odot$. Distance is 1 kpc. }
\label{paramm}
\end{figure}

The effect of ejected mass is shown in figure \ref{paramm} for a CO (left) and 
an ONe (right) nova. In both cases, the effect of ejected mass at early epochs 
is opposite to that at later times. At early times, 
the larger the ejected mass the larger the opacity of the envelope to gamma-rays; 
therefore, novae with larger ejected masses emit smaller fluxes. On the contrary,
at later times (after $\sim$1 day, see figure \ref{paramm}), the envelope is more 
transparent and the fluxes almost directly reflect the amount of $^{18}$F (i.e., 
the larger the ejected mass the larger the flux). In the ONe novae, the fluxes 
emitted in the ``plateau'' phase directly reflect the amount of $^{22}$Na (thus 
the ratio of fluxes equals the ratio of ejected masses). 

\begin{figure}
\setlength{\unitlength}{1cm}
\begin{picture}(18,8)
\put(0,-2){\makebox(9,10){\epsfxsize=8cm \epsfbox{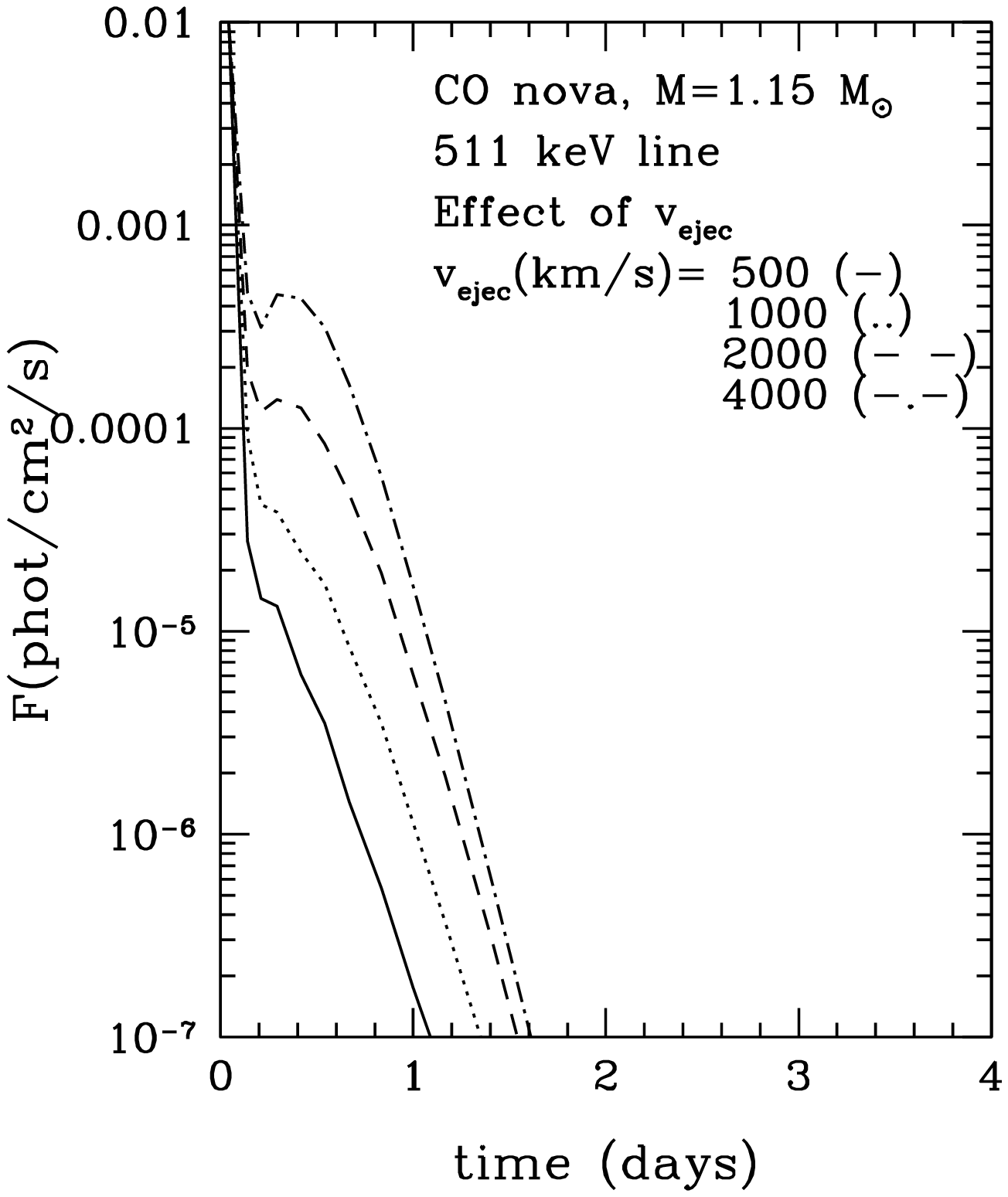}}}
\put(7.5,-2){\makebox(9,10){\epsfxsize=8cm \epsfbox{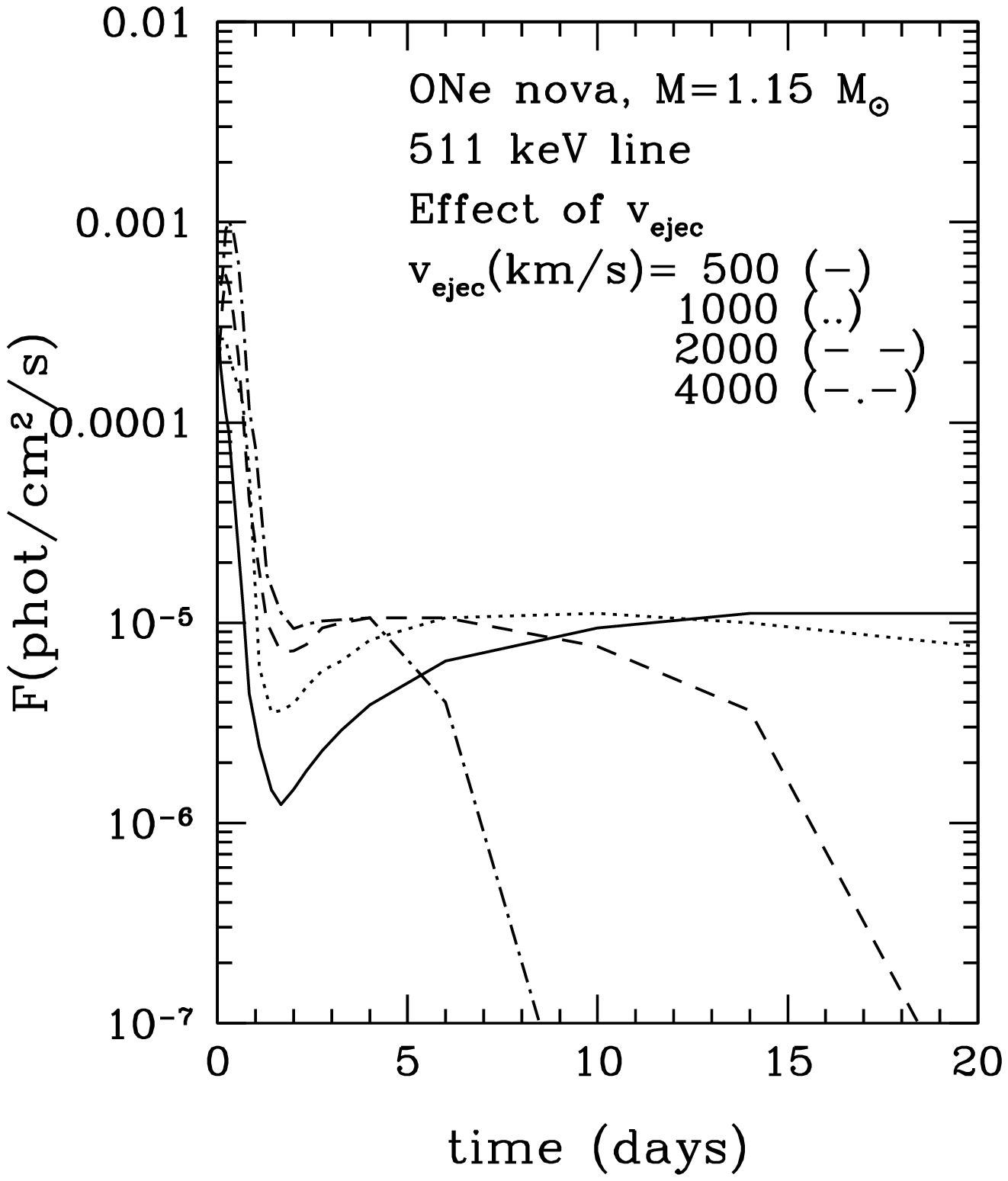}}}
\end{picture}
\caption{(Left) Light curves for the 511 keV line for a CO nova of 
1.15 M$_\odot$, for a range of parametrized velocities of the ejecta. The 
value indicated corresponds to the outermost shell. 
(Right) Same for an ONe nova of 1.15 M$_\odot$. Distance is 1 kpc.}
\label{paramv}
\end{figure}

The influence of the velocity of the ejecta is displayed in figure \ref{paramv}. 
At early epochs, the gamma-ray flux increases with envelope velocities 
because of increased transparency. Later on (around 1 day after peak 
temperature), ONe novae differ from CO novae, because of their much larger 
$^{22}$Na content: the decline of the gamma-ray flux in ONe novae is delayed  by 
$^{22}$Na positrons annihilating in the envelope, until the moment when 
the envelope 
becomes transparent to positrons and also these annihilations fade away. This 
happens some weeks later, the exact figure depending on the 
transparency of the envelope; therefore, the smaller the 
velocity, the smaller the transparency at a given time and the longer the duration 
of this phase (see figure \ref{paramv} right). 

\vspace{-0.5cm}
\section{Discussion and prospects for detectability}
\vspace{-0.5cm}
We have shown that the continuum (20-511 keV) and 511 keV line light curves 
provide a direct insight into the dynamics of the expanding envelope, as well as 
information about the nova type and its $^{18}$F (CO and ONe novae) and $^{22}$Na 
(ONe novae) content. The fluxes in various continuum bands and in the 511 keV 
line are more intense than those in the 478 keV line and 1275 keV lines, but 
their short duration and early appearence makes them
impossible to detect with standard pointed observations. Only all-sky monitors, 
able to do surveys in the range from 20 to 600 keV, are suited for such type of 
detection.
Attempts have been made with the TGRS instrument onboard the WIND satellite 
\citep{Har99}, the BATSE onboard the Compton Gamma-Ray Observatory 
\citep{Fis91,Her00}. The negative results obtained up to now are consitent with 
our predicted fluxes. Therefore, we have to wait for more sensitive instruments 
(EXIST, Advanced COMPTON Telescope), or hopefully the shield of the future 
INTEGRAL spectrometer SPI \citep{Jea99}, to confirm the theoretical predictions
and to extract all the crucial information about the nova phenomenon that only 
the gamma-rays can provide.

This research has been partially supported by the CICYT-P.N.I.E.
(ESP98-1348) and the DGES (PB98-1183-C03-02; PB98-1183-C03-03)


\begin{table*}
\begin{center}
\caption{Radioactivities in novae ejecta relevant for prompt emission 
($^{13}$N and $^{18}$F at 1h after T$_{max}$)}
\label{radioac}
\vspace{0.3cm}
\renewcommand{\arraystretch}{1.5}
\begin{tabular}{ccccccc}
\hline
Nova        & M$_{\rm wd}(\rm M_\odot)$ & M$_{\rm ejec}(\rm M_\odot)$  
            & $^{13}$N (M$_\odot$)      & $^{18}$F (M$_\odot$)        
            & $^{22}$Na (M$_\odot$)\\
\hline
CO          & 0.8                       & 6.2x10$^{-5}$                  
            & 1.5x10$^{-7}$             & 1.8x10$^{-9}$                
            & 7.4x10$^{-11}$\\
CO          & 1.15                      & 1.3x10$^{-5}$                  
            & 2.3x10$^{-8}$             & 2.6x10$^{-9}$                  
            & 1.1x10$^{-11}$\\
ONe         & 1.15                      & 2.6x10$^{-5}$                  
            & 2.9x10$^{-8}$             & 5.9x10$^{-9}$                  
            & 6.4x10$^{-9}$\\
ONe         & 1.25                      & 1.8x10$^{-5}$                  
            & 3.8x10$^{-8}$             & 4.5x10$^{-9}$                  
            & 5.9x10$^{-9}$\\
\hline
\end{tabular}
\end{center}
\end{table*}

\end{document}